\documentclass[aps,prd,floats,floatfix,superscriptaddress,nofootinbib,preprintnumbers]{revtex4-1}

\usepackage{graphicx}
\usepackage{epstopdf}
\usepackage{hyperref}
\usepackage[utf8]{inputenc} 
\usepackage{bbold}
\usepackage{color}
\usepackage{booktabs} 
\usepackage{array} 
\usepackage{paralist} 
\usepackage{verbatim} 
\usepackage{subfig} 
\usepackage{amssymb}
\usepackage{amsmath}
\usepackage{accents}

\newcommand{\bi}{\begin{itemize}}
\newcommand{\ei}{\end{itemize}}
\newcommand{\ben}{\begin{enumerate}}
\newcommand{\een}{\end{enumerate}}
\newcommand{\be}{\begin{equation}}
\newcommand{\ee}{\end{equation}}
\newcommand{\bea}{\begin{eqnarray}}
\newcommand{\eea}{\end{eqnarray}}
\newcommand{\ba}{\begin{align}}
\newcommand{\ea}{\end{align}}
\newcommand{\nn}{\nonumber}
\newcommand*{\pbar}[1]{\accentset{\left(-\right)\vspace{-.5mm}}{#1}}

\def\tev{{\rm TeV}}
\def\mgamc{{\sc MadGraph5\_aMC@NLO} }
\def\GM{Georgi-Machacek }

\begin{document}

\title{Automatic predictions in the Georgi-Machacek model \\ at next-to-leading order accuracy}
\author{C\'eline Degrande}
\email{celine.degrande@durham.ac.uk}
\affiliation{Institute for Particle Physics Phenomenology, Department of Physics, Durham University, Durham DH1 3LE, United Kingdom}
\author{Katy Hartling} 
\email{khally@physics.carleton.ca}
\affiliation{Ottawa-Carleton Institute for Physics, Carleton University, 1125 Colonel By Drive, Ottawa, Ontario K1S 5B6, Canada}
\author{Heather~E.~Logan}
\email{logan@physics.carleton.ca}
\affiliation{Ottawa-Carleton Institute for Physics, Carleton University, 1125 Colonel By Drive, Ottawa, Ontario K1S 5B6, Canada}
\author{Andrea D.~Peterson}
\email{apeterso@physics.carleton.ca}
\affiliation{Ottawa-Carleton Institute for Physics, Carleton University, 1125 Colonel By Drive, Ottawa, Ontario K1S 5B6, Canada}
\author{Marco~Zaro} 
\email{zaro@lpthe.jussieu.fr}
\affiliation{Sorbonne Universit\'{e}s, UPMC Univ.~Paris 06, UMR 7589, LPTHE, F-75005, Paris, France}
\affiliation{CNRS, UMR 7589, LPTHE, F-75005, Paris, France}

\preprint{IPPP/15/72, DCPT/15/144, MCnet-15-34}

\date{December 3, 2015}                                           

\begin{abstract}
We study the phenomenology of the Georgi-Machacek model at next-to-leading order (NLO) in QCD matched to parton shower, using a fully-automated tool chain based on {\sc MadGraph5\_aMC@NLO}
and {\sc FeynRules}. We focus on the production of the fermiophobic custodial fiveplet scalars $H_5^0$, $H_5^\pm$, and $H_5^{\pm\pm}$ through vector boson fusion (VBF), associated production with a vector boson ($VH_5$), and scalar pair production ($H_5H_5$). For these production mechanisms we compute NLO corrections 
to production rates as well as to differential distributions. Our results demonstrate that the Standard Model (SM) overall $K$-factors for such processes cannot in general be directly applied to beyond-the-SM distributions, due both to differences in the scalar electroweak charges and to variation of the $K$-factors over the differential distributions.  
\end{abstract}

\maketitle 

\section{Introduction}

A deeper understanding of the scalar sector is a primary objective of the CERN Large Hadron Collider (LHC).  In addition to precisely measuring the 125 GeV Higgs boson, Run II of the LHC will dedicate its efforts to searching for signs of additional Higgs particles, which arise in a number of beyond-the-Standard-Model (BSM) scenarios. One such scenario is the \GM (GM) model~\cite{Georgi:1985nv,Chanowitz:1985ug}, which extends the Standard Model (SM) with two scalar isospin triplets in a way that preserves the SM value of $\rho = M_W^2/M_Z^2 \cos^2{\theta_W} =1$ at tree level. The phenomenology of the GM model has previously been studied in Refs.~\cite{Gunion:1990dt,Gunion:1989ci,Akeroyd:1998zr,Haber:1999zh,Aoki:2007ah,Godfrey:2010qb,Low:2010jp,Logan:2010en,Carmi:2012in,Chang:2012gn,Chiang:2012cn,Kanemura:2013mc,Englert:2013zpa,Killick:2013mya,Belanger:2013xza,Englert:2013wga,Chiang:2013rua,Efrati:2014uta,Hartling:2014zca,Chiang:2014hia,Chiang:2014bia,Godunov:2014waa,Hartling:2014aga,Chiang:2015kka,Godunov:2015lea,Chiang:2015rva}, including the application of a variety of constraints upon the model parameter space. It has been shown to possess a decoupling limit, and can thus accommodate an SM-like 125 GeV boson~\cite{Hartling:2014zca}. Furthermore, the tree-level couplings of this SM-like Higgs to fermions and vector bosons may be enhanced in comparison to the SM~\cite{Hartling:2014aga}, a feature that cannot be accommodated in models that contain only scalars in $SU(2)$ singlet or doublet representations. The GM model can also be embedded in more elaborate theoretical scenarios, such as little Higgs \cite{Chang:2003un,Chang:2003zn} and supersymmetric \cite{Cort:2013foa,Garcia-Pepin:2014yfa,Delgado:2015bwa} models, or generalized to larger $SU(2)$ multiplets \cite{Logan:2015xpa}.

The \GM model provides a useful benchmark framework for BSM Higgs searches. In addition to an SM-like scalar singlet $h$, the GM model also contains an extra scalar singlet $H$, a triplet $H_3$, and a fiveplet $H_5$ under the custodial symmetry. The structure of the model with respect to the custodial singlet and triplet states is similar to that of the two Higgs doublet model (2HDM); as a result, the experimental searches and extensive analysis for 2HDM states can often be recast in terms of the GM singlet and triplet scalars~\cite{Hartling:2014aga}. 
It is therefore particularly interesting to focus on the custodial fiveplet states, $H_5^0$, $H_5^\pm$ and $H_5^{\pm\pm}$. These scalars are fermiophobic and couple preferentially to vector bosons. As a result, the GM fiveplet contains two features that are absent from both the SM and the 2HDM: a doubly charged scalar $H_5^{\pm\pm}$ and charged scalar states that couple to vector bosons. Consequently, the fermiophobic fiveplet states are produced primarily through the vector boson fusion (VBF) and associated production ($VH_5$) modes. This is in contrast to the 2HDM, where the heavy scalars are dominantly produced through associated production with a top quark or in top decays. These features lead to unique phenomenology and can be used to parametrize effects not captured by other common benchmark models. 

For the \GM model to be truly useful as an LHC benchmark, efficient and accurate calculations must be accessible to both phenomenologists and experimentalists. Great strides have been made in reducing both theoretical and experimental uncertainties, making next-to-leading order (NLO) or higher order calculations standard practice. Therefore, we describe the use of a fully-automated tool chain (which combines the {\sc FeynRules}~\cite{Alloul:2013bka} and \mgamc~\cite{Alwall:2014hca} frameworks with the calculator {\sc GMCALC}~\cite{Hartling:2014xma}) to produce NLO differential distributions in the GM model, focusing on the examples of VBF, $VH_5$, and $H_5 H_5$ production of the fiveplet states. In particular, we illustrate the insufficiency of extending the SM overall $K$-factors to BSM distributions, due to two factors. First, differential $K$-factors can vary substantially for certain distributions (particularly in the case of VBF). Second, the overall $K$-factors for differently-charged states can be somewhat different. These considerations are important for accurately determining the effects of typical selection cuts, which is essential for measuring new states in the event of a discovery.  

This paper is organized as follows. In the following section, we describe in more detail the scalar potential, spectrum, and couplings of the \GM model. In Sec.~\ref{sec:framework}, we then outline the tools used for our fully-automated NLO calculations. Finally, in Secs.~\ref{sec:VBF}, \ref{sec:VH} and \ref{sec:HH}, we present cross sections, $K$-factors, and differential distributions for VBF, $VH_5$, and pair production ($H_5H_5$), respectively, of the fiveplet states.  We conclude in Sec.~\ref{sec:conclusions}.  For completeness, some details of the scalar potential of the GM model are collected in an appendix.  The model files for the automated tool chain used to produce these results are publicly available on \url{http://feynrules.irmp.ucl.ac.be/wiki/GeorgiMachacekModel}.

\section{The model} 

The scalar sector of the GM model~\cite{Georgi:1985nv,Chanowitz:1985ug} consists of the usual complex isospin doublet $(\phi^+,\phi^0)$ with hypercharge\footnote{We normalize the hypercharge operator such that $Q = T^3 + Y/2$.} $Y = 1$, a real triplet $(\xi^+,\xi^0,\xi^-)$ with $Y = 0$, and  a complex triplet $(\chi^{++},\chi^+,\chi^0)$ with $Y=2$.  The doublet is responsible for the fermion masses as in the SM.

The scalar potential is chosen by hand to preserve a global $SU(2)_L \times SU(2)_R$ symmetry.  This ensures that $\rho = M_W^2/M_Z^2\cos^2\theta_W = 1$ at tree level, as required by precise experimental measurements~\cite{PDG}.  In order to make the global $SU(2)_L \times SU(2)_R$ symmetry explicit, we write the doublet in the form of a bidoublet $\Phi$ and combine the triplets to form a bitriplet $X$:
\begin{equation}
	\Phi = \left( \begin{array}{cc}
	\phi^{0*} &\phi^+  \\
	-\phi^{+*} & \phi^0  \end{array} \right), \qquad
	X =
	\left(
	\begin{array}{ccc}
	\chi^{0*} & \xi^+ & \chi^{++} \\
	 -\chi^{+*} & \xi^{0} & \chi^+ \\
	 \chi^{++*} & -\xi^{+*} & \chi^0  
	\end{array}
	\right).
	\label{eq:PX}
\end{equation}
The vacuum expectation values (vevs) are defined by $\langle \Phi  \rangle = \frac{ v_{\phi}}{\sqrt{2}} I_{2\times2}$  and $\langle X \rangle = v_{\chi} I_{3 \times 3}$, where $I$ is the unit matrix and the Fermi constant $G_F$ fixes the combination of vevs,
\begin{equation}
	v_{\phi}^2 + 8 v_{\chi}^2 \equiv v^2 = \frac{1}{\sqrt{2} G_F} \approx (246~{\rm GeV})^2.
	\label{eq:vevrelation}
\end{equation} 
These vevs are parametrized in terms of a mixing angle $\theta_H$ according to
\begin{equation}
	c_H \equiv \cos\theta_H = \frac{v_{\phi}}{v}, \qquad \qquad
	s_H \equiv \sin\theta_H = \frac{2\sqrt{2}\,v_\chi}{v}.
\end{equation}
The quantity $s_H^2$ represents the fraction of the squared gauge boson masses $M_W^2$ and $M_Z^2$ that is generated by the vev of the triplets, while $c_H^2$ represents the fraction generated by the usual Higgs doublet. The most general scalar potential that preserves the custodial $SU(2)$ symmetry may be found in Appendix~\ref{A:GMmodel}.

After symmetry breaking, the physical fields can be organized by their transformation properties under the custodial $SU(2)$ symmetry into a fiveplet, a triplet, and two singlets.  The fiveplet and triplet states are given by
\begin{eqnarray}
	&&H_5^{++} = \chi^{++}, \qquad \quad
	H_5^+ = \frac{\left(\chi^+ - \xi^+\right)}{\sqrt{2}}, \qquad \quad
	H_5^0 = \sqrt{\frac{2}{3}} \xi^0 - \sqrt{\frac{1}{3}} \chi^{0,r}, \nonumber \\
	&&H_3^+ = - s_H \phi^+ + c_H \frac{\left(\chi^++\xi^+\right)}{\sqrt{2}}, \qquad \quad
	H_3^0 = - s_H \phi^{0,i} + c_H \chi^{0,i},
\end{eqnarray}
where we have decomposed the neutral fields into real and imaginary parts according to
\begin{equation}
	\phi^0 \to \frac{v_{\phi}}{\sqrt{2}} + \frac{\phi^{0,r} + i \phi^{0,i}}{\sqrt{2}},
	\qquad \quad
	\chi^0 \to v_{\chi} + \frac{\chi^{0,r} + i \chi^{0,i}}{\sqrt{2}}, 
	\qquad \quad
	\xi^0 \to v_{\chi} + \xi^0.
	\label{eq:decomposition}
\end{equation}
The states of the custodial fiveplet $(H_5^{\pm \pm}, H_5^{\pm}, H_5^0)$ have a common mass $m_5$ and the states of the custodial triplet $(H_3^{\pm}, H_3^0)$ have a common mass $m_3$.  Because the states in the custodial fiveplet contain no doublet field content, they do not couple to fermions (i.e. they are fermiophobic).

The two custodial singlets mix by an angle $\alpha$, and the resulting mass eigenstates are given by
\begin{equation}
	h = \cos \alpha \, \phi^{0,r} - \sin \alpha \, H_1^{0\prime},  \qquad \quad
	H = \sin \alpha \, \phi^{0,r} + \cos \alpha \, H_1^{0\prime},
	\label{mh-mH}
\end{equation}
where 
\begin{equation}
	H_1^{0 \prime} = \sqrt{\frac{1}{3}} \xi^0 + \sqrt{\frac{2}{3}} \chi^{0,r}.
\end{equation}
We denote their masses by $m_h$ and $m_H$.  The singlet $h$ is normally identified as the 125~GeV SM-like Higgs boson discovered at the LHC~\cite{Aad:2012tfa,Chatrchyan:2012xdj,Aad:2015zhl}. Formulae for the masses $m_h$, $m_H$, $m_3$, and $m_5$, as well as the mixing angle $\alpha$, may be found in Appendix~\ref{A:GMmodel}.

The fiveplet states couple to vector bosons according to the following Feynman rules~\cite{HHG,Godfrey:2010qb,Hartling:2014zca}:
\begin{eqnarray}
	H_5^0 W^+_{\mu} W^-_{\nu} : &&\quad
	\sqrt{\frac{2}{3}} i g^2 v_{\chi} g_{\mu\nu}
	= 2 (\sqrt{2} G_F)^{1/2} M_W^2 \left( -\frac{1}{\sqrt{3}} s_H \right) (-i g_{\mu\nu}),
                \label{eq:hwpwmvert}
	\\
	H_5^0 Z_{\mu} Z_{\nu} : &&\quad
	- \sqrt{\frac{8}{3}} i \frac{g^2}{c_W^2} v_{\chi} g_{\mu\nu}
	= 2 (\sqrt{2} G_F)^{1/2} M_Z^2 \left( \frac{2}{\sqrt{3}} s_H \right) (-i g_{\mu\nu}),
                \label{eq:hzzvert}
        \\
	H_5^+ W^-_{\mu} Z_{\nu} : &&\quad -\sqrt{2} i \frac{g^2}{c_W} v_{\chi} g_{\mu\nu}
		= 2 ( \sqrt{2} G_F )^{1/2} M_W M_Z ( s_H ) (-i g_{\mu\nu}),
                \label{eq:hwzvert}
        \\
	H_5^{++} W^-_{\mu} W^-_{\nu} :&& \quad
	2 i g^2 v_{\chi} g_{\mu\nu} 
	= 2 (\sqrt{2} G_F)^{1/2} M_W^2 \left( -\sqrt{2} s_H \right) (-i g_{\mu\nu}),
                \label{eq:hwpwpvert}
\end{eqnarray}
where we write the coupling in multiple forms to make contact with the notation of Refs.~\cite{Godfrey:2010qb,Bolzoni:2011cu}.  The triplet vev $v_{\chi}$ is called $v^{\prime}$ in Ref.~\cite{Godfrey:2010qb}, and the 
factors $F_{VV}$ in Eq.~(5.2) of Ref.~\cite{Bolzoni:2011cu} correspond in this model to  
\begin{eqnarray}
    F_{W^+W^-}=-\frac{1}{\sqrt 3}s_H &&\qquad (H_5^0\textrm{ production}), \\
    F_{ZZ}=\frac{2}{\sqrt 3} s_H &&\qquad (H_5^0\textrm{ production}), \\
    F_{W^\pm Z}=s_H &&\qquad (H_5^\pm\textrm{ production}), \\
    F_{W^\pm W^\pm}=-\sqrt{2} s_H &&\qquad (H_5^{\pm\pm}\textrm{ production}).
\end{eqnarray}
Note in particular that, for $H_5^0$, one cannot simply rescale the vector boson fusion cross section of the SM Higgs boson because the ratio of the $WW$ and $ZZ$ couplings is different than in the SM.

Additionally, two fiveplet scalars may also couple to a single vector boson through the following interactions:\footnote{As we consider only the fiveplet states in this work, we quote only the relevant interactions involving $H_5$  scalar states and gauge bosons. A full set of Feynman rules for the GM scalar couplings may be found in Ref.~\cite{Hartling:2014zca}.}
\begin{eqnarray}
\gamma_\mu H_5^{+} H_5^{+*}:&&\quad i e (p_{+} - p_{+*})_\mu, \\
\gamma_\mu H_5^{++} H_5^{++*}:&&\quad 2 i e (p_{++} - p_{++*})_\mu, \\
Z_\mu H_5^{+} H_5^{+*}:&&\quad \frac{i e}{2 s_W c_W}(1-2s_W^2)(p_{+} - p_{+*})_\mu, \\
Z_\mu H_5^{++} H_5^{++*}:&&\quad \frac{i e}{s_W c_W}(1-2s_W^2)(p_{++} - p_{++*})_\mu, \\
W^+_\mu H_5^{+*} H_5^0: &&\quad \frac{\sqrt{3} i e}{2 s_W}(p_{+*} - p_{0})_\mu, \\
W^+_\mu H_5^{+} H_5^{++*}: &&\quad \frac{ie}{\sqrt{2} s_W}(p_{+} - p_{++*})_\mu, 
\end{eqnarray}
where all fields are incoming and in each case $p_Q$ is the incoming momentum of the scalar with charge $Q$.  Note that these are independent of the mixing angle $s_H$.

There are theoretical constraints on the \GM model from considerations of perturbativity and vacuum stability~\cite{Aoki:2007ah,Chiang:2012cn,Hartling:2014zca}, as well as indirect experimental constraints from the measurements of oblique parameters ($S$, $T$, $U$), $Z$-pole observables ($R_b$), and $B$-meson observables~\cite{Haber:1999zh,Kanemura:2013mc,Englert:2013zpa,Chiang:2012cn,Chiang:2013rua,Hartling:2014aga}. Currently the strongest of the indirect experimental bounds arises from measurements of $b \to s \gamma$, which constrain the triplet vev $v_\chi \leq 65$~GeV ($s_H \leq 0.75$)~\cite{Hartling:2014aga}.  Additionally, the ATLAS like-sign $WWjj$ cross-section measurement, reinterpreted in the context of the GM model in Ref.~\cite{Chiang:2014bia}, excludes a doubly-charged Higgs $H_5^{\pm\pm}$ with masses in the range $140\leq m_5 \leq 400$~GeV at $s_H = 0.5$, and $100\leq m_5 \leq 700$ GeV at $s_H=1$, under the assumption of a 100\% branching fraction for $H_5^{++} \to W^+W^+$. An ATLAS search for singly charged scalars in the VBF production channel similarly excludes $240\leq m_5\leq 700$ GeV for $s_H=1$ under the assumption of a 100\% branching fraction for $H_5^+\rightarrow W^+Z$~\cite{Aad:2015nfa}. Additional constraints on $v_\chi$ as a function of the BSM Higgs masses have been obtained in Ref.~\cite{Chiang:2015kka} using ATLAS data from several search channels.

For the simulations that follow, we consider a single benchmark point in the GM model, generated using the calculator {\sc GMCALC} \cite{Hartling:2014xma}.\footnote{Our benchmark point corresponds to the default point in {\sc GMCALC}. The choice of masses, mixing angles, and $M_{1,2}$ as input parameters corresponds to the {\sc GMCALC} input set 3.} This point is allowed by all the constraints discussed above.  We use the following values for the scalar masses, mixing angles, and additional parameters $M_{1,2}$ as inputs:
\begin{align}
&m_h = 125~\text{GeV}, & \sin\alpha  = -0.303, \nn \\
&m_H = 288~\text{GeV}, & \sin{\theta_H} = 0.194, \nn \\
&m_3 =  304~\text{GeV}, & M_1  =  100~\text{GeV}, \nn \\
&m_5 = 340~\text{GeV}, & M_2  =  100~\text{GeV}.
\label{modelparams}
\end{align}
The parameters $M_{1,2}$ are dimensionful parameters in the scalar potential [see Eq.~(\ref{eq:potential})] that affect the values of the couplings between scalars. The corresponding values for the underlying parameters of the scalar potential are given in Appendix~\ref{A:GMmodel}. While we specify the complete parameter set, note that all the $H_5 V V$ couplings are proportional to $s_H$. Therefore, both the VBF and $VH_5$ production cross sections of the $H_5$ states depend only on two parameters, $s_H$ and $m_5$, and the $H_5H_5$ production cross sections depend only on $m_5$.  At this parameter point the total widths of the $H_5$ states are about 0.3~GeV; therefore in our simulations we will take the final-state $H_5$ particle(s) to be produced on shell.

Finally, we choose the following set of SM inputs: 
\begin{align}
M_W &= 80.399  \text{ GeV}, &
M_Z &=  91.188  \text{ GeV}, \nn \\
\Gamma_W &= 2.085  \text{ GeV},  &
\Gamma_Z &=  2.495  \text{ GeV}, \nn \\
 G_F&=  1.166 \times10^{-5}  \text{ GeV}^{-2}. 
\label{smparams}
\end{align}
$\alpha_{EM} = 1/132.35$ is computed at tree level from $M_W$, $M_Z$, and $G_F$.

\section{Computational framework\label{sec:framework}}
In this work we take advantage of a fully automated framework developed to study the phenomenology of BSM processes at NLO accuracy in
QCD, including the matching to parton shower (PS). The framework is based on \mgamc~\cite{Alwall:2014hca}. In order to generate a code
capable of computing NLO corrections to a BSM process, some extra information has to be
provided besides the usual tree-level Feynman rules. This extra information involves the ultraviolet (UV) renormalization counterterms and a subset of the rational terms that are
needed in the numerical reduction of virtual matrix elements (which are normally referred to as the $R_2$ 
terms)~\cite{Ossola:2008xq}. The calculation of the UV and $R_2$ terms starting from the model Lagrangian has been automatized via
the {\sc NLOCT} package~\cite{Degrande:2014vpa}, based on {\sc FeynRules}~\cite{Alloul:2013bka} and 
{\sc FeynArts}~\cite{Hahn:2000kx}. Once the UV and $R_2$ Feynman rules have been generated, they are exported together with the tree-level Feynman rules as a {\sc Python}\ module in the Universal FeynRules Output ({\sc UFO})\ format~\cite{Degrande:2011ua}. The {\sc Python}\ module
can be loaded by any matrix-element generator, such as \mgamc\!. When the code for the process is written, the {\sc UFO}\ information
is translated into helicity routines~\cite{Murayama:1992gi} by {\sc ALOHA}~\cite{deAquino:2011ub}.
\mgamc is a meta-code that automatically generates the code to perform the simulation of any process up to NLO accuracy in QCD. The
simulation can be performed either at fixed order or by generating event samples which can be passed to PS. The automation
of the NLO QCD corrections has been achieved by exploiting the {\sc FKS}~\cite{Frixione:1995ms, Frixione:1997np} subtraction 
scheme to subtract the infrared singularities of real-emission matrix elements, as automated in {\sc MadFKS}~\cite{Frederix:2009yq}. Loops are computed
by {\sc MadLoop}~\cite{Hirschi:2011pa}, which exploits the OPP~\cite{Ossola:2006us} method as well as Tensor Integral 
Reduction~\cite{Passarino:1978jh, Davydychev:1991va}; these are implemented in  {\sc CutTools}~\cite{Ossola:2007ax} 
and {\sc IREGI}~\cite{iregi} respectively, which are supplemented by an in-house implementation of {\sc OpenLoops}~\cite{Cascioli:2011va}. 
Finally, the event generation and matching to PS is done following the 
{\sc MC@NLO}\ procedure~\cite{Frixione:2002ik}. Matching to {\sc Herwig6}~\cite{Corcella:2000bw}, 
{\sc Pythia6}~\cite{Sjostrand:2006za},\footnote{Ordered in virtuality or in transverse momentum, 
with the latter only for processes with no light partons in the final state.}
{\sc Herwig++}~\cite{Bahr:2008pv}, and {\sc Pythia8}~\cite{Sjostrand:2007gs} is available.

As a consequence, the only input needed to simulate processes in the GM model is the implementation of the model in {\sc FeynRules}.
We have validated our framework by comparing total cross sections at NLO for VBF with the results of the {\sc VBF@NNLO} 
code~\cite{Bolzoni:2010xr, Zaro:2010fc, Bolzoni:2011cu} and found agreement within the integration uncertainties. 

\section{VBF Production}
\label{sec:VBF}

In the SM, VBF production has been calculated to a rather high level of accuracy: QCD corrections are known up to next-to-next-to-leading order (NNLO) for the total cross 
section~\cite{Han:1992hr,Bolzoni:2010xr,Bolzoni:2011cu} and for differential observables at the parton level~\cite{Figy:2003nv,Arnold:2008rz,Cacciari:2015jma}.  The QCD corrections to the fully inclusive cross sections are fairly moderate, at the level of a few percent. However, the corrections to differential observables are more significant, with NNLO corrections reaching 5--10\% relative to the NLO rate.  At both the inclusive and differential levels, the computation of NNLO corrections relies on the so-called structure-function approach~\cite{Han:1992hr}, which neglects color- and kinematically-suppressed contributions~\cite{vanNeerven:1984ak,Blumlein:1992eh,Figy:2007kv,Harlander:2008xn,Bolzoni:2011cu} arising, for example, from the exchange of gluons between the two quark lines. Results at NLO in QCD including parton shower matching have been computed in Refs.~\cite{Nason:2009ai,Frixione:2013mta}, where it has been found that the typical effect of the shower is to improve the description of jet-related observables by including the effect of extra radiation. NLO electroweak (EW) corrections are also known~\cite{Ciccolini:2007jr,Ciccolini:2007ec} and are found to be comparable in size to the NLO QCD ones.

The situation is less satisfactory for BSM scenarios like the Georgi-Machacek model. Although the total cross section can be computed up to NNLO 
accuracy in QCD~\cite{Zaro:2010fc,Bolzoni:2010xr,Zaro:2015ika}, no fully differential prediction exists beyond leading order (LO). As seen in the SM case, corrections to the inclusive total cross sections do not fully capture the behavior at the differential level. In this section we aim to improve this situation, by presenting for the first time fully differential results at NLO in QCD including matching to the parton shower.

\subsection{\label{sec:VBFsim}Simulation}

The code for VBF production of a fiveplet state in the GM model can be generated and executed in \mgamc\ with the commands
\begin{verbatim}
> import model GM_UFO
> generate p p > H5p j j $$ w+ w- z [QCD]
> output VBF_h5p_NLO
> launch
\end{verbatim}
Note that  we veto $W$ and $Z$ bosons in the
$s$-channel with the \texttt{\$\$} syntax. The example above generates the code for $H_5^+$ production. For the other states, 
$H_5^{--}$, $H_5^{-}$, $H_5^{0}$, $H_5^{++}$, the code can be generated by replacing the \texttt{H5p} label with \texttt{H5pp\~}, \texttt{H5p\~},
\texttt{H5z}, \texttt{H5pp} respectively.

We present results for VBF in the GM model at the LHC Run II energy ($\sqrt{s}=13~\tev$) at LO and NLO accuracy,
in both cases matched to {\sc Pythia8}. We use the NNPDF 2.3 LO1 and NLO parton density function (PDF) sets~\cite{Ball:2012cx} consistently with the order 
of the computation.
We keep the renormalization and factorization scales fixed to the $W$ boson mass, as the typical transverse momentum of the 
tagging jets is of the same order of magnitude. To obtain the uncertainty due to scale variations, 
we vary the renormalization and factorization scales independently in the range
\be
M_W/2 \le \mu_R, \mu_F \le 2 M_W.
\ee
We recall that the computation of scale and PDF uncertainties in \mgamc can be performed without the need of extra runs using the reweighting technique presented in Ref.~\cite{Frederix:2011ss}.
We employ {\sc FastJet} \cite{Cacciari:2005hq,Cacciari:2011ma} to cluster hadrons into jets, using the anti-$k_T$ algorithm~\cite{Cacciari:2008gp} with a radius parameter $\Delta R=0.4$. 
A minimum jet $p_T$ of 30 GeV is required. 

In addition, we consider the effect of typical selection cuts used in VBF analyses. These {\it VBF cuts} require that there are at least two jets, and that the two hardest jets satisfy the conditions
\bea
 y_j &<& 4.5,  \nn \\
 |y_{j_1}-y_{j_2}| &> &4.0,  \nn \\
 m(j_1,j_2) &>& 600 \text{ GeV,}
 \label{eq:VBFcuts}
\eea
where $y_j$ is the jet rapidity and $m(j_1,j_2)$ is the invariant mass of the two jets.

\subsection{Results}
\label{sec:VBFres}

In Tables~\ref{tab:VBFxs} and \ref{tab:VBFxscut} we present the cross sections at the inclusive level and with the VBF cuts of Eq.~(\ref{eq:VBFcuts}),
respectively, for the production via VBF of each of the fiveplet states. Results are shown at LO+PS and NLO+PS, together
with the fractional uncertainties obtained from scale variations. First, we note that the $K$ factors without and with cuts are rather similar to each other.  Furthermore, the $K$ factors for the different fiveplet states are also rather similar, and lie around 1.1. The production of more negatively-charged Higgs bosons receives slightly larger QCD corrections; this effect, related to the cross section's sensitivity to valence versus sea quarks, becomes slightly more pronounced when VBF cuts are applied. The inclusion of NLO corrections also has the effect of reducing
the scale uncertainties to the 1--2\% level. The different dependence on the initial state quarks of the various processes is also reflected in the efficiency of the VBF cuts. The fraction of events that survives the VBF cuts (tabulated under ``cut efficiency'' in Table~\ref{tab:VBFxscut}) varies from 44\% in the case of $H_5^{--}$ production to 47\% in the case of $H_5^{++}$ production, and is essentially unaffected by inclusion of the NLO corrections.

\begin{table}[htp]
\begin{center}
\begin{tabular}{  c  c  c  c  }
    \hline
     Process & LO (fb) & NLO (fb) & $K$  \\ \hline\hline
        $pp \rightarrow H^{--}_5 jj$  & $14.94^{+5.4\%}_{ -5.1\%} $ & $16.72^{+1.4\%}_{ -0.7\%}$ & 1.12\\
        $pp \rightarrow H^-_5 jj$  & $16.94^{+5.3\%}_{ -5.0\%} $ & $18.66^{+1.3\%}_{ -0.5\%}$ & 1.10\\
        $pp \rightarrow H^0_5 jj$  & $21.08^{+5.4\%}_{ -5.0\%} $ & $22.89^{+1.4\%}_{ -0.5\%}$ & 1.09\\
        $pp \rightarrow H^+_5 jj$& $28.07^{+5.8\%}_{ -5.3\%} $   & $30.14^{+1.5\%}_{ -0.7\%} $ & 1.07\\ 
        $pp \rightarrow H^{++}_5 jj$  & $40.90^{+6.6\%}_{ -5.9\%} $ & $43.56^{+1.4\%}_{ -0.6\%}$ & 1.07\\
    \hline
\end{tabular}
\end{center}
\caption{Cross sections and $K$-factors for $H_5^{\pm}$ VBF production, with scale uncertainties.}
\label{tab:VBFxs}
\end{table}

\begin{table}[htp]
\begin{center}
\begin{tabular}{  c  c  c  c  c }
    \hline
     Process & LO (fb) & NLO (fb) & $K$ & Cut efficiency \\ \hline\hline
        $pp \rightarrow H^{--}_5jj$    & $6.58^{+7.1\%}_{ -6.5\%}$  & $7.47^{+1.3\%}_{ -1.2\%}$ & 1.13 & 0.44 \\
        $pp \rightarrow H^-_5jj$   & $7.75^{+7.0\%}_{ -6.4\%}$ & $8.66^{+1.2\%}_{ -0.9\%}$ &1.12 & 0.46 \\
        $pp \rightarrow H^0_5jj$   & $9.82^{+7.1\%}_{ -6.4\%}$   & $10.71^{+1.2\%}_{ -0.7\%}$ & 1.09 & 0.47 \\
        $pp \rightarrow H^+_5jj$  & $13.29^{+7.3\%}_{ -6.5\%}$    & $14.26^{+1.4\%}_{ -0.8\%}$ &1.07 & 0.47 \\
        $pp \rightarrow H^{++}_5jj$  & $19.36^{+7.9\%}_{ -7.0\%}$  & $20.49^{+1.2\%}_{ -0.6\%}$ & 1.06  & 0.47 \\
    \hline
\end{tabular}
\end{center}
\caption{Cross sections and $K$-factors for $H_5^{\pm}$ VBF production, with scale uncertainties,  after applying the VBF cuts given in Eq.~(\ref{eq:VBFcuts}).  Also shown is the fraction of NLO events that survive the VBF cuts (``cut efficiency'').}
\label{tab:VBFxscut}
\end{table}

\begin{figure}
\resizebox{0.4\textwidth}{!}{\includegraphics{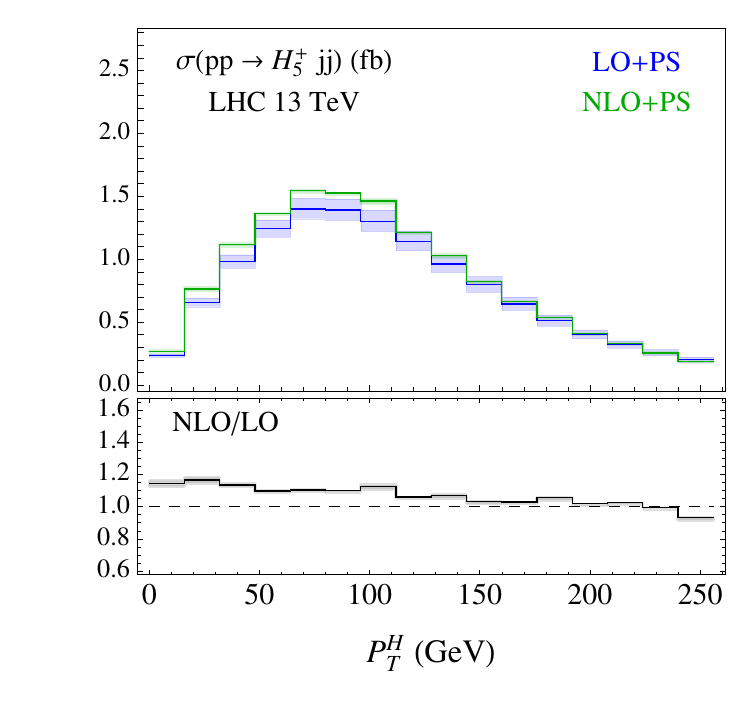}}
\resizebox{0.4\textwidth}{!}{\includegraphics{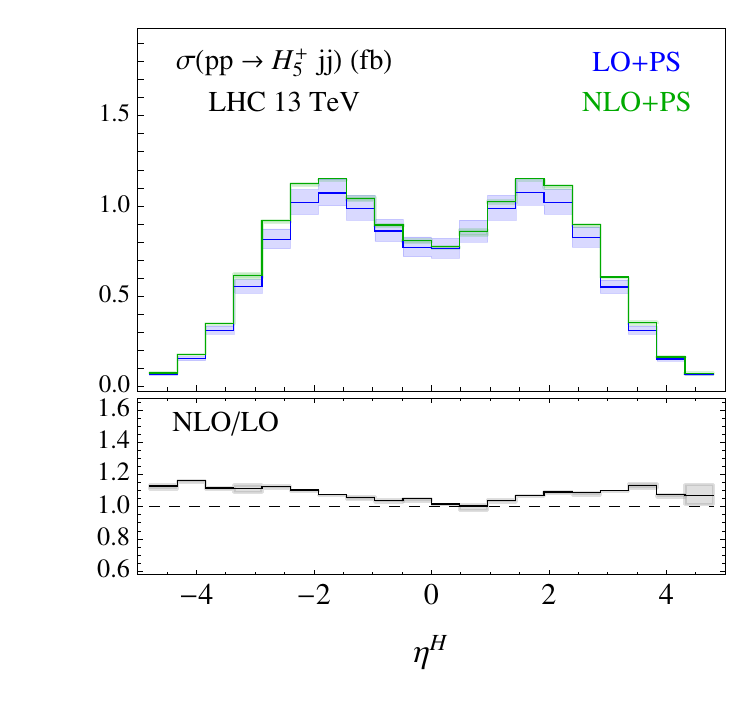}}
\resizebox{0.4\textwidth}{!}{\includegraphics{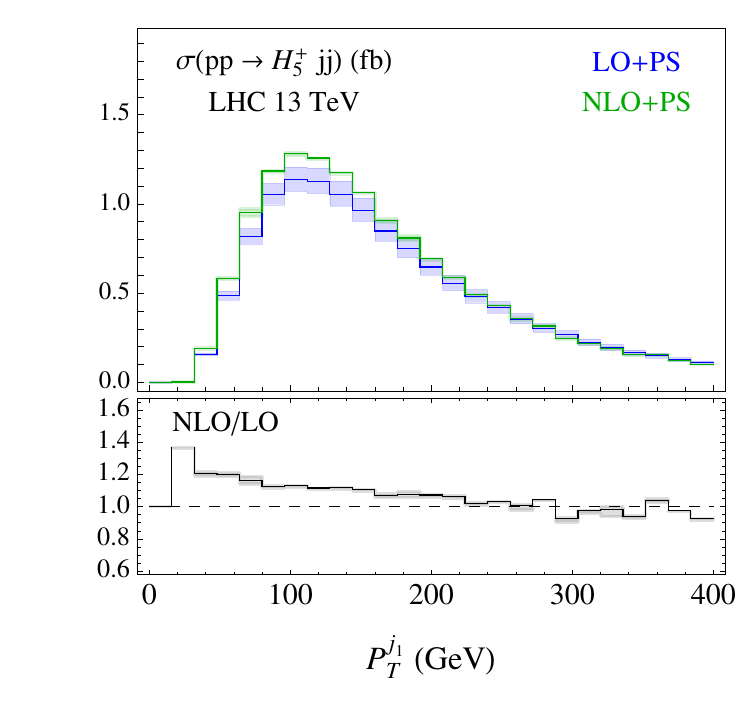}}
\resizebox{0.4\textwidth}{!}{\includegraphics{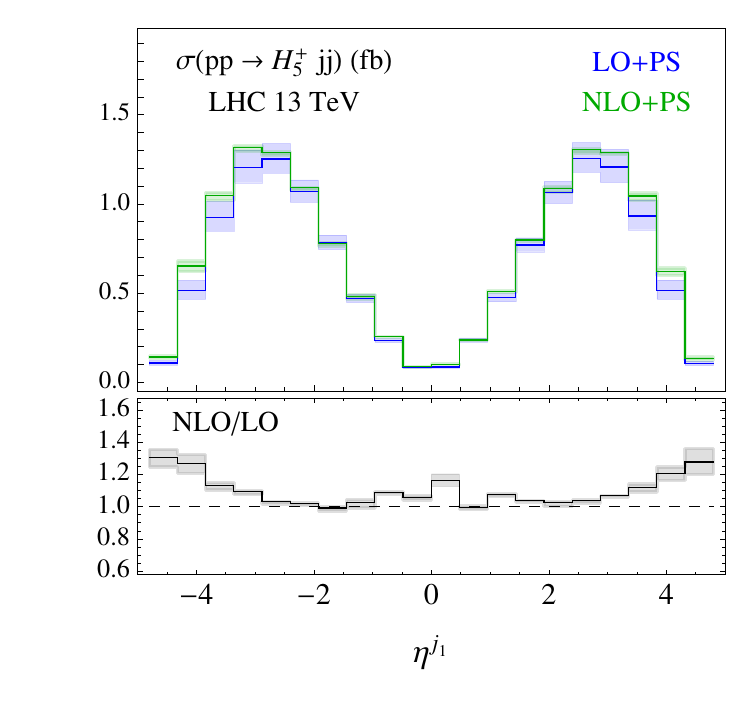}}
\resizebox{0.4\textwidth}{!}{\includegraphics{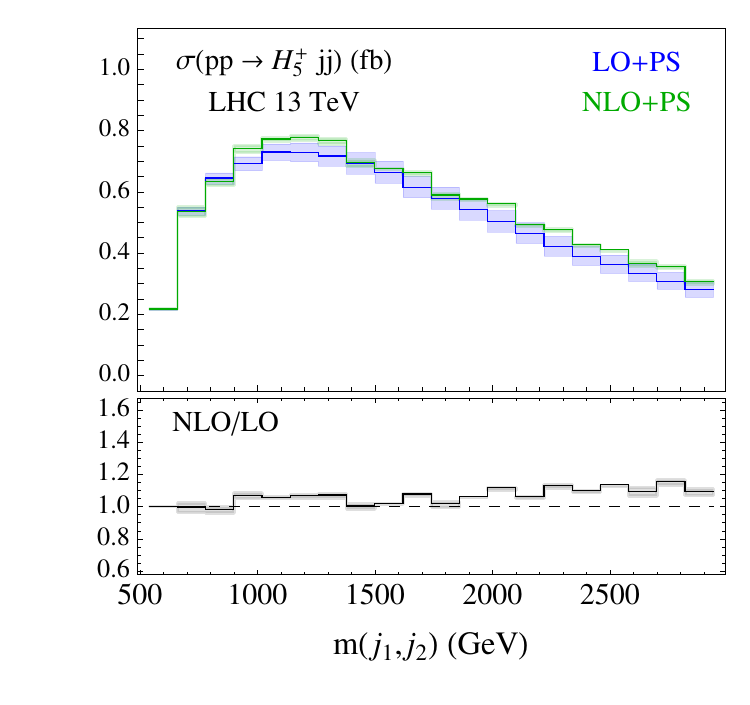}}
\resizebox{0.4\textwidth}{!}{\includegraphics{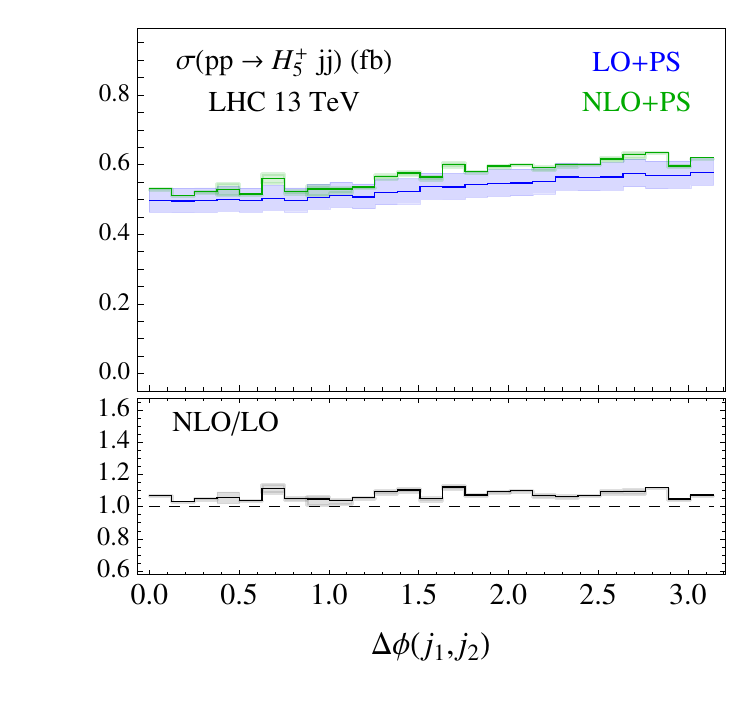}}

\caption{Differential distributions for VBF production of the $H_5^+$ boson, with the VBF cuts of  Eq.~(\ref{eq:VBFcuts}) (see text for details). The distributions for other $H_5$ states are very similar, differing primarily in overall normalization.}
\label{fig:vbf_h5p}
\end{figure}

We turn now to study the effect of NLO corrections on differential observables, focusing on the representative case of $H_5^+$ production in VBF.
In Figure \ref{fig:vbf_h5p}, we show the LO+PS and NLO+PS distributions for a number of observables. In particular we consider the transverse momentum $p_T$ and 
pseudorapidity $\eta$ of the Higgs boson ($H_5$) and of the hardest jet ($j_1$), as well as the invariant mass $m(j_1,j_2)$ and azimuthal separation $\Delta \phi(j_1,j_2)$ of the two hardest jets.  The shaded bands show the scale uncertainties at both LO and NLO.
The VBF cuts of Eq.~(\ref{eq:VBFcuts}) have been applied. For each observable, we also show in the inset 
the differential $K$-factor: that is, the bin-by-bin ratio of the NLO prediction over the LO central value, with the shaded band reflecting the NLO scale uncertainty. As in the case of SM VBF Higgs boson production, the $K$-factor is in
general not constant over the differential distributions. This effect is most visible for the hardest-jet observables. Therefore, a fully-differential computation at NLO+PS is strongly preferable to ensure realistic signal simulations.

\section{$VH_5$ Production}
\label{sec:VH}

We now consider the associated production of a GM fiveplet state together with a $W^{\pm}$  or $Z$ boson. In the SM, the associated production of a Higgs boson with a vector boson is known to NNLO in QCD for the total cross section~\cite{Kniehl:1990iva,Han:1991ia,Ohnemus:1992bd,Baer:1992vx,Brein:2003wg,Brein:2011vx,Altenkamp:2012sx,Harlander:2014wda}; the two-loop corrections increase the inclusive cross section by less than 5\% at the LHC~\cite{Brein:2003wg}. The QCD corrections to the differential observables are also known to NNLO~\cite{Ferrera:2011bk,Ferrera:2014lca}, leading to increases of 5--20\% in comparison with the NLO results. N$^3$LO threshold corrections of about 0.1\% have also been calculated in Ref.~\cite{Kumar:2014uwa}. These results have been implemented along with the electroweak corrections~\cite{Ciccolini:2003jy,Brein:2004ue} in the {\sc vh$@$nnlo} code~\cite{Brein:2012ne}.

In following sections, we present rates and distributions for $VH_5$ production at NLO for the Georgi-Machacek model. 

\subsection{Simulation}
The code for associated production of a fiveplet state in the GM model (in this example $H_5^+$) and a SM vector boson ($W^-$ or $Z$, decaying leptonically with $l = e$ or $\mu$)
can be generated and executed in \mgamc\ with the commands 
\begin{verbatim}
> import model GM_UFO
> add process p p > H5p l- vl~ [QCD]
> add process p p > H5p l+ l- [QCD]
> output VH_h5p_NLO
> launch
\end{verbatim}
In this case, we include the leptonic decay of the gauge bosons at the matrix-element level, so that spin correlations and 
off-shell effects are automatically taken into account. As in the VBF case, the extension to the other states in the Higgs fiveplet is straightforward.
We set the renormalisation and factorization 
scales to the invariant mass of the (reconstructed) $VH_5$ system, $\mu_R = \mu_F = M_{VH}$.  

We consider two sets of cuts.  In the first case, we require only basic cuts on leptons and missing
transverse energy. Leptons are required to satisfy the transverse momentum and pseudorapidity cuts
\be
p_T^l > 30\text{ GeV\quad and\quad } \eta_l < 2.5. 
 \label{eq:lepid}
\ee
For $WH_5$ associated production, we also cut on the transverse missing energy, reconstructed from neutrinos in the event record:
\be
	E_T^{\rm miss} > 30~{\rm GeV}.
 \label{eq:met}
\ee

In the second case, we consider a boosted regime, which is often used to enhance the signal-to-background ratio in SM $VH$ 
searches~\cite{Butterworth:2008iy, ATLAS:2009elr}, 
by requiring the following additional cuts on the Higgs
and the reconstructed gauge bosons' transverse momenta:      
\be   p_T^H> 200\text{ GeV\quad and\quad }p_T^V > 190\text{ GeV,} \label{eq:VHcut} \ee
as suggested in \cite{Heinemeyer:2013tqa}.

\subsection{Results}
    In Tables \ref{tab:VHxs} and  \ref{tab:VHxscut}, we show the cross sections for $VH_5$ production of $H_5$ states at LO+PS and NLO+PS with basic cuts and with the additional boosted-regime cuts, respectively. The cross sections include the leptonic branching fractions of the gauge bosons. Note that $Z H_5^{\pm\pm}$ production of the doubly charged states is forbidden by charge conservation. We find that the $K$-factors are larger than for VBF and, similar to the SM case, lie around $1.3$. Furthermore,
    the value of $K$-factors without and with the boosted-regime cuts of Eq.~(\ref{eq:VHcut}) are essentially identical. We notice that processes with a more 
    negatively-charged final state (which are therefore more sensitive to sea quarks) have slightly larger $K$-factors. As in the case of VBF, processes with a more positively-charged final state have a larger fraction of events which survive the cuts.

\begin{table}
\begin{center}
\begin{tabular}{  c  c  c  c  }
    \hline
     Process & LO (fb) & NLO (fb) & $K$  \\ \hline\hline
        $pp \rightarrow H^-_5 l^+ l^-$   & $0.01701^{+7.0\%}_{ -6.4\%}$ & $0.02294^{+1.2\%}_{ -0.9\%}$ &1.35 \\
        $pp \rightarrow H^0_5\  l^+ l^-$   & $0.03570^{+7.1\%}_{ -6.4\%}$   & $0.04736^{+1.2\%}_{ -0.7\%}$ & 1.33\\
        $pp \rightarrow H^+_5 l^+ l^-$  & $0.03338^{+7.3\%}_{ -6.5\%}$    & $0.04332^{+1.4\%}_{ -0.8\%}$ &1.30 \\
	\hline
        $pp \rightarrow H^{--}_5 l^+ \nu_l$    & $0.10852^{+7.1\%}_{ -6.5\%}$  & $0.14668^{+1.3\%}_{ -1.2\%}$ & 1.35\\
        $pp \rightarrow H^-_5\  l^+ \nu_l$   & $0.08573^{+7.0\%}_{ -6.4\%}$ & $0.11394^{+1.2\%}_{ -0.9\%}$ &1.33 \\
        $pp \rightarrow H^0_5\  l^\pm\ \pbar{\nu}_l$   & $0.05354^{+7.1\%}_{ -6.4\%}$   & $0.07053^{+1.2\%}_{ -0.7\%}$ & 1.32\\
        $pp \rightarrow H^+_5\  l^-  \bar{\nu}_l$  & $0.08438^{+7.3\%}_{ -6.5\%}$    & $0.11192^{+1.4\%}_{ -0.8\%}$ &1.33 \\
        $pp \rightarrow H^{++}_5 l^- \bar{\nu}_l$  & $0.21096^{+7.9\%}_{ -7.0\%}$  & $0.27332^{+1.2\%}_{ -0.6\%}$ & 1.30 \\
      \hline
\end{tabular}
\end{center}
\caption{Cross sections and $K$-factors for $V H_5$ production after the basic lepton identification cuts given in Eqs.~(\ref{eq:lepid}) and (\ref{eq:met}), with scale uncertainties.  For the first three processes the Higgs is produced in association with a $Z$ boson, and for the remainder with a $W$ boson.}
\label{tab:VHxs}
\end{table}

\begin{table}
\begin{center}
\begin{tabular}{  c  c  c  c  c }
    \hline
     Process & LO (fb) & NLO (fb) & $K$ & Cut efficiency \\ \hline\hline
        $pp \rightarrow H^-_5 l^+ l^-$   & $0.00741^{+7.0\%}_{ -6.4\%}$ & $0.00989^{+1.2\%}_{ -0.9\%}$ &1.34 & 0.43\\
        $pp \rightarrow H^0_5\  l^+ l^-$   & $0.01601^{+7.1\%}_{ -6.4\%}$   & $0.02112^{+1.2\%}_{ -0.7\%}$ & 1.32 & 0.45\\
        $pp \rightarrow H^+_5 l^+ l^-$  & $0.01549^{+7.3\%}_{ -6.5\%}$    & $0.02011^{+1.4\%}_{ -0.8\%}$ &1.30 & 0.46\\
	\hline
        $pp \rightarrow H^{--}_5 l^+ \nu_l$    & $0.04803^{+7.1\%}_{ -6.5\%}$  & $0.06515^{+1.3\%}_{ -1.2\%}$ & 1.36 & 0.44\\
        $pp \rightarrow H^-_5 l^+ \nu_l$   & $0.03921^{+7.0\%}_{ -6.4\%}$ & $0.05188^{+1.2\%}_{ -0.9\%}$ &1.32 & 0.46\\
        $pp \rightarrow H^0_5\  l^\pm\ \pbar{\nu}_l$   & $0.02497^{+7.1\%}_{ -6.4\%}$   & $0.03278^{+1.2\%}_{ -0.7\%}$ & 1.31 & 0.46\\
        $pp \rightarrow  H^+_5 l^- \bar{\nu}_l$  & $0.03897^{+7.3\%}_{ -6.5\%}$    & $0.05163^{+1.4\%}_{ -0.8\%}$ &1.32 & 0.46 \\
        $pp \rightarrow H^{++}_5 l^- \bar{\nu}_l$  & $0.10158^{+7.9\%}_{ -7.0\%}$  & $0.13148^{+1.2\%}_{ -0.6\%}$ & 1.29 &0.48 \\
    \hline
\end{tabular}
\end{center}
\caption{Cross sections and $K$-factors for $V H_5$ production after applying the additional boosted-regime cuts given in Eq.~(\ref{eq:VHcut}). Also shown is the fraction of NLO events that survive the boosted-regime cuts (``cut efficiency'').}
\label{tab:VHxscut}
\end{table}

\begin{figure}
\begin{center}
\resizebox{0.4\textwidth}{!}{\includegraphics{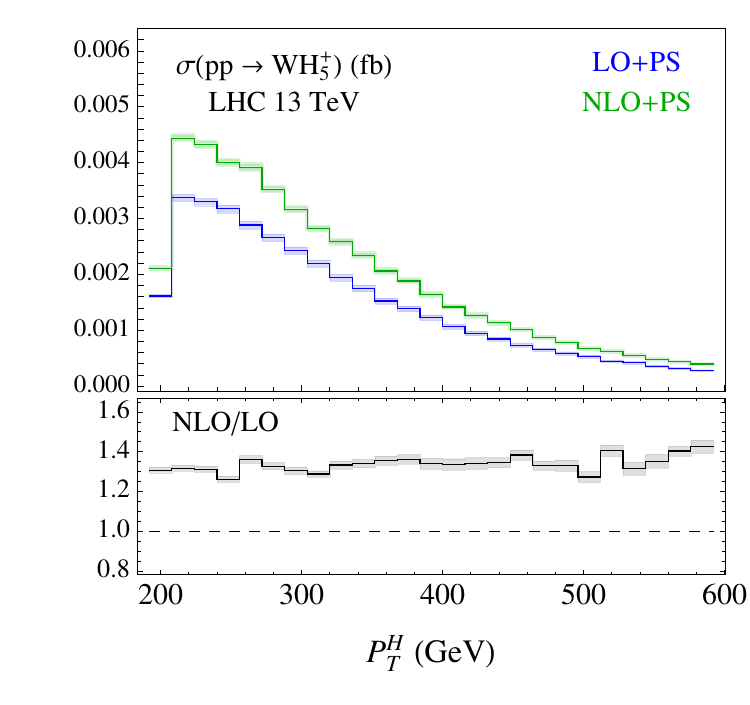}}
\resizebox{0.4\textwidth}{!}{\includegraphics{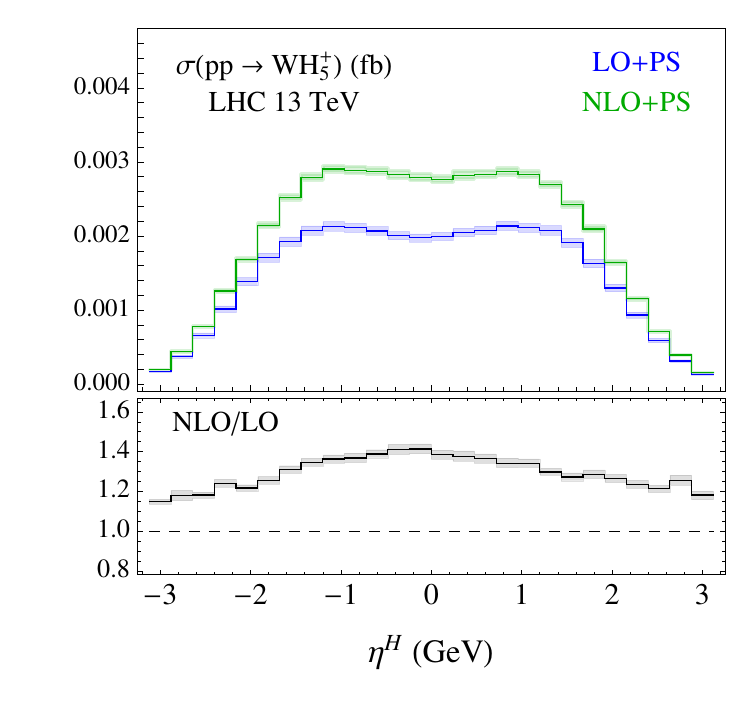}}
\resizebox{0.4\textwidth}{!}{\includegraphics{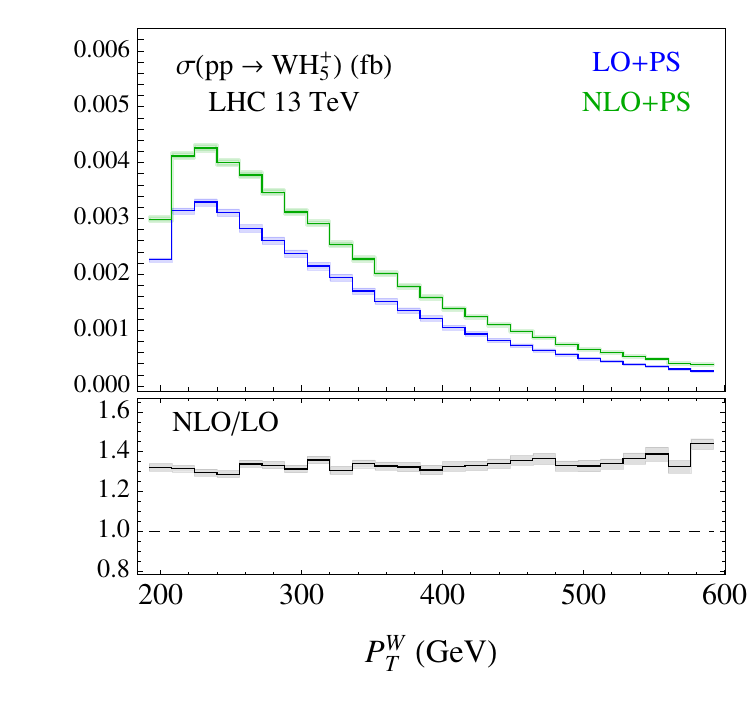}}
\resizebox{0.4\textwidth}{!}{\includegraphics{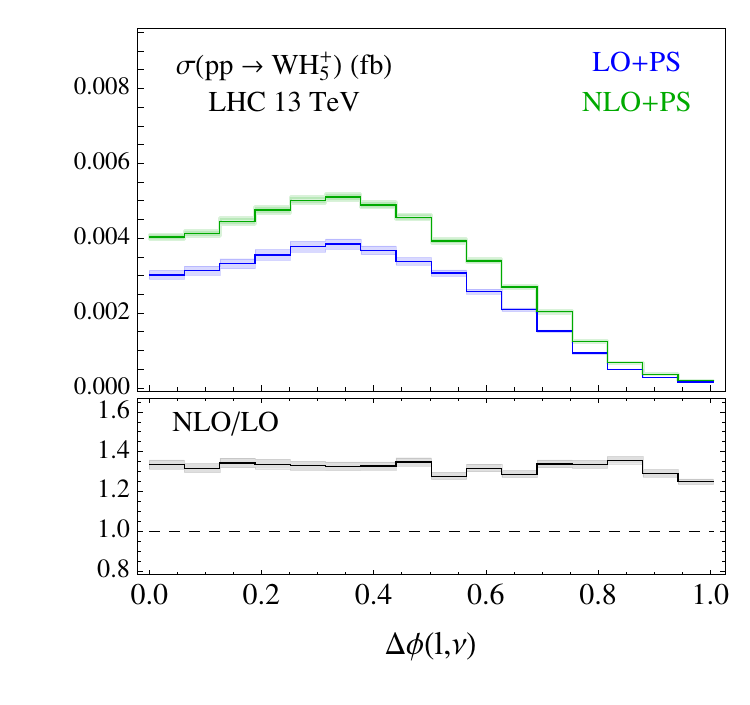}}

\caption{Differential distributions for $W H_5^+$ associated production, with the cuts in  Eq.~\ref{eq:VHcut}. The distributions for other $H_5$ states or for associated production with a $Z$ boson are very similar, differing primarily in overall normalization.}
\label{fig:wh_h5p}
\end{center}
\end{figure}

In Figure \ref{fig:wh_h5p}, we present the LO+PS and NLO+PS distributions and $K$-factors for $W^- H_5^+$ production under the boosted-regime $VH_5$ cuts given in Eq.~(\ref{eq:VHcut}); the distributions for $ZH_5^+$ production are similar in shape. We show the transverse momentum $p_T$ and pseudorapidity $\eta$ of the Higgs, the
transverse momentum of the reconstructed vector boson (using monte carlo truth information) and the azimuthal separation $\Delta \phi$ between the lepton 
and the neutrino. 
In this case we find that the differential $K$-factors are generally constant over the distributions considered, with the exception of the Higgs pseudorapidity; in this case the $K$-factor has a maximum of around 1.4 in the central region, which reduces to a minimum of around 1.2 for
a Higgs produced in the forward or backward regions.

\section{$H_5H_5$ Production\label{sec:HH}}

Finally we consider double Higgs production of two $H_5$ states in the GM model.  In contrast to the SM, pair production of the fiveplet scalars is generally dominated by Drell-Yan-like processes.\footnote{In the SM, Higgs pair production is dominated by gluon fusion at the LHC. The rate of this production mode is known to be quite small, and receives important QCD corrections at NLO~\cite{Plehn:1996wb,Dawson:1998py,Binoth:2006ym,Baglio:2012np}. Corrections to the inclusive cross section have been obtained at NNLO~\cite{deFlorian:2013uza,deFlorian:2013jea,Grigo:2014jma}, while corrections to differential observables are known at NLO~\cite{Frederix:2014hta}. The NNLO corrections to the inclusive cross section are quite large, on the order of 20\%~\cite{deFlorian:2013jea} in comparison to the NLO result at 14 TeV. The effect of dimension-6 operators arising from new physics has also been considered at NLO in Ref.~\cite{Grober:2015cwa}, which found that the new couplings could alter the $K$-factors relevant to SM-like Higgs pair production by a few percent.} 
The exception is $H_5^0 H_5^0$ pair production (which we therefore do not consider below), as there is no $Z H_5^0 H_5^0$ vertex due to the same symmetry considerations that forbid the $ZHH$ coupling in the SM. $H_5^0 H_5^0$ pairs could be produced through VBF, and the $H_5^0 H_5^0$, $H_5^+ H_5^-$, and $H_5^{++} H_5^{--}$ final states could also be produced via gluon fusion through an off-shell $h$ or $H$. These processes have very small cross sections and are not considered here. 

\subsection{Simulation}
The code for the pair production of two fiveplet states in the GM model, for example $H_5^{--} H_5^+$,
can be generated and executed in \mgamc\ with the commands 
\begin{verbatim}
> import model GM_UFO
> generate p p > H5pp~ H5p [QCD]
> output H5mm_H5p_NLO
> launch
\end{verbatim}
Once again, the extension to the other combinations of states in the Higgs fiveplet is straightforward.
We set the renormalisation and factorization 
scales to the invariant mass of the Higgs pair, $\mu_R = \mu_F = M_{HH}$. We do not consider additional cuts on these processes. 

\subsection{Results}

\begin{figure}
\begin{center}
\resizebox{0.4\textwidth}{!}{\includegraphics{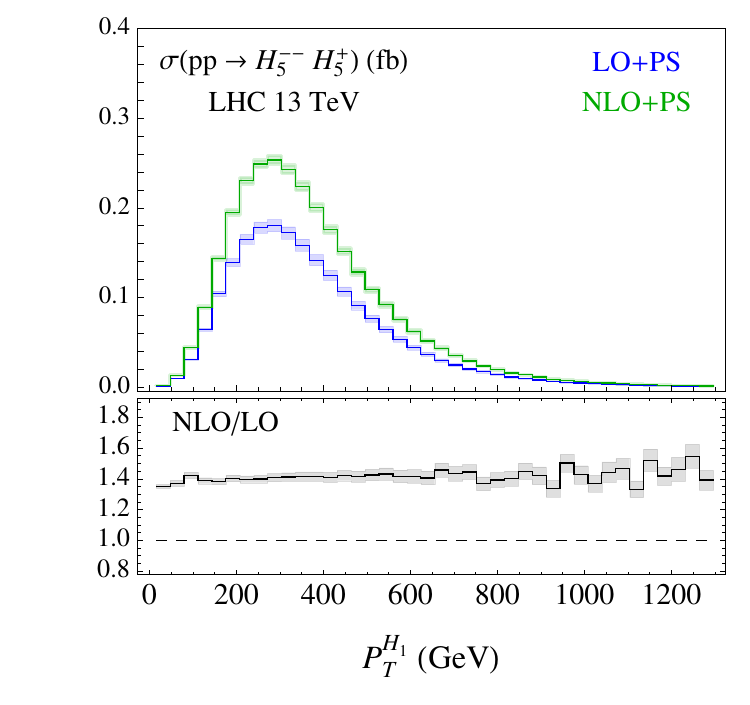}}
\resizebox{0.4\textwidth}{!}{\includegraphics{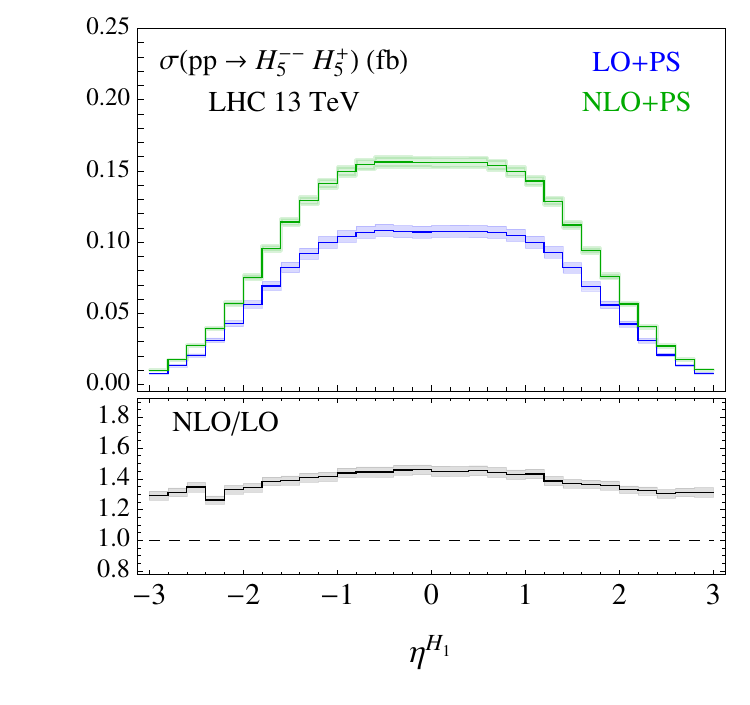}}
\resizebox{0.4\textwidth}{!}{\includegraphics{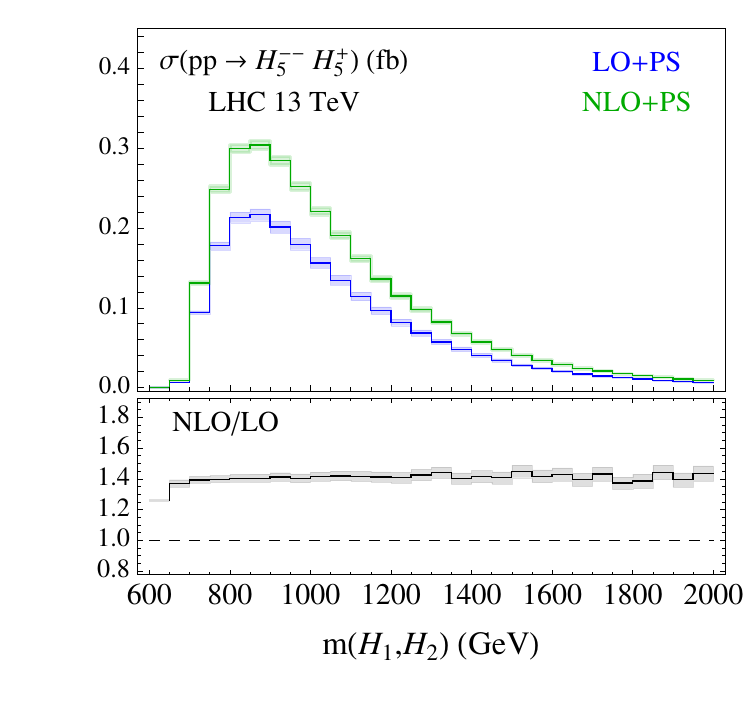}}

\caption{Differential distributions for $H_5^{--} H_5^+$ associated production. The distributions for other $H_5$ states are very similar, differing primarily in overall normalization.}
\label{fig:hh_mmp}
\end{center}
\end{figure}

In Table~\ref{tab:HHxs} we show the cross sections for $H_5H_5$ production at LO+PS and NLO+PS, without cuts. In Figure \ref{fig:hh_mmp}, we show the LO+PS and NLO+PS distributions and $K$-factors for $H_5^{--}H_5^+$ production. We show the transverse momentum $p_T$ and pseudorapidity $\eta$ of the scalar $H_5^{--}$, and the invariant mass of the two scalars. The $p_T$ and $\eta$ distributions of the other scalar $H_5^+$ are similar. 

As in the case of $VH_5$ production, the differential $K$-factors are generally constant over the distributions considered, with the exception of the Higgs pseudorapidities; in this case the $K$-factor has a maximum slightly above 1.4 in the central region, which reduces to roughly 1.3 for a Higgs produced in the forward or backward regions.

\begin{table}
\begin{center}
\begin{tabular}{  c  c  c  c  }
    \hline
     Process & LO (fb) & NLO (fb) & $K$  \\ \hline\hline
        $pp \rightarrow H^{--}_5 H^{+}_5 $    & $2.113^{+4.4\%}_{ -4.2\%}$  & $2.977^{+2.2\%}_{ -2.1\%}$ & 1.41\\
        $pp \rightarrow H^{-}_5 H^{0}_5$   & $3.174^{+4.4\%}_{ -4.2\%}$ & $4.464^{+2.2\%}_{ -2.1\%}$ &1.41 \\
        $pp \rightarrow H^{--}_5 H^{++}_5$   & $7.589^{+4.4\%}_{ -4.2\%}$   & $10.499^{+2.2\%}_{ -2.1\%}$ & 1.38\\
        $pp \rightarrow H^{-}_5 H^{+}_5$  & $1.897^{+4.4\%}_{ -4.2\%}$    & $2.624^{+2.2\%}_{ -2.1\%}$ &1.38 \\
        $pp \rightarrow H^{0}_5 H^{+}_5$  & $7.128^{+4.6\%}_{ -4.4\%}$  & $9.671^{+2.2\%}_{ -2.2\%}$ & 1.36 \\
	$pp \rightarrow H^{-}_5 H^{++}_5$  & $4.752^{+4.6\%}_{ -4.4\%}$  & $6.448^{+2.2\%}_{ -2.2\%}$ & 1.36 \\
      \hline
\end{tabular}
\end{center}
\caption{Cross sections and $K$-factors for $H_5 H_5$ production, with scale uncertainties.  The first two processes proceed through an $s$-channel $W^-$, the next two through a $Z$ and the last two through a $W^+$.}
\label{tab:HHxs}
\end{table}

\section{Conclusions}
\label{sec:conclusions}

We have presented cross sections, differential distributions, and $K$-factors for the production of fermiophobic fiveplet scalars in the Georgi-Machacek model at NLO accuracy in QCD, including the matching to parton showers. We considered production through VBF, $VH_5$, and $H_5H_5$ associated production at the benchmark point of Eq.~(\ref{modelparams}). Our results demonstrate the importance of a fully differential simulation at NLO+PS in order to accurately simulate the signal at the LHC. Automated tools make such a simulation possible with a very limited effort. 
For what concerns $VH_5$ and $H_5H_5$ production, the description of the final state can be further improved by including the effect of
the radiation of extra jets at NLO accuracy, for example using the Fx-Fx~\cite{Frederix:2012ps} or UNLOPS~\cite{Lonnblad:2012ix} merging technique, which are both automatized within \mgamc.  
The model files for the automated tool chain used to produce these results are publicly available on \url{http://feynrules.irmp.ucl.ac.be/wiki/GeorgiMachacekModel}.

\begin{acknowledgments}
We thank K. Kumar for helpful discussions about the Georgi-Machacek model and {\sc FeynRules} and we are grateful to F. Maltoni and L. Barak for having 
encouraged us to pursue the present project. C.D. is a Durham International Junior Research Fellow
 and has been supported in part by the Research Executive Agency of the European Union
under Grant Agreement PITN-GA-2012-315877 (MC-Net). K.H., H.E.L., and A.P.\ were supported by the Natural Sciences and Engineering Research Council of Canada.  K.H.\ was also supported by the Government of Ontario through an Ontario Graduate Scholarship. M.Z.\ is supported by the European Union's Horizon 2020 research and innovation
programme under the Marie Sklodovska-Curie grant agreement No 660171 and in part by the ERC grant
Higgs@LHC and by the ILP LABEX (ANR-10-LABX-63), in turn supported by French state funds
managed by the ANR within the ``Investissements d'Avenir'' programme under
reference ANR-11-IDEX-0004-02. 
\end{acknowledgments}

\clearpage
\appendix

\section{The scalar potential and masses of the Georgi-Machacek model \label{A:GMmodel}}

The most general gauge-invariant scalar potential involving these fields that conserves custodial $SU(2)$ can be written as\footnote{A translation table to other parametrizations in the literature has been given in an Appendix of Ref.~\cite{Hartling:2014zca}. Note that Refs.~\cite{Gunion:1990dt,Gunion:1989ci,Akeroyd:1998zr,Haber:1999zh,Godfrey:2010qb,Carmi:2012in,Chang:2012gn,Englert:2013zpa,Englert:2013wga,Efrati:2014uta,Belanger:2013xza} impose an additional $Z_2$ symmetry on this potential, such that $M_1=M_2=0$ and the model has no decoupling limit.}~\cite{Hartling:2014zca}  
\begin{eqnarray}
	V(\Phi,X) &= & \frac{\mu_2^2}{2}  \text{Tr}(\Phi^\dagger \Phi) 
	+  \frac{\mu_3^2}{2}  \text{Tr}(X^\dagger X)  
	+ \lambda_1 [\text{Tr}(\Phi^\dagger \Phi)]^2  
	+ \lambda_2 \text{Tr}(\Phi^\dagger \Phi) \text{Tr}(X^\dagger X)   \nonumber \\
          & & + \lambda_3 \text{Tr}(X^\dagger X X^\dagger X)  
          + \lambda_4 [\text{Tr}(X^\dagger X)]^2 
           - \lambda_5 \text{Tr}( \Phi^\dagger \tau^a \Phi \tau^b) \text{Tr}( X^\dagger t^a X t^b) 
           \nonumber \\
           & & - M_1 \text{Tr}(\Phi^\dagger \tau^a \Phi \tau^b)(U X U^\dagger)_{ab}  
           -  M_2 \text{Tr}(X^\dagger t^a X t^b)(U X U^\dagger)_{ab}.
           \label{eq:potential}
\end{eqnarray} 
Here the $SU(2)$ generators for the doublet representation are $\tau^a = \sigma^a/2$ with $\sigma^a$ being the Pauli matrices, the generators for the triplet representation are
\begin{equation}
	t^1= \frac{1}{\sqrt{2}} \left( \begin{array}{ccc}
	 0 & 1  & 0  \\
	  1 & 0  & 1  \\
	  0 & 1  & 0 \end{array} \right), \qquad  
	  t^2= \frac{1}{\sqrt{2}} \left( \begin{array}{ccc}
	 0 & -i  & 0  \\
	  i & 0  & -i  \\
	  0 & i  & 0 \end{array} \right), \qquad 
	t^3= \left( \begin{array}{ccc}
	 1 & 0  & 0  \\
	  0 & 0  & 0  \\
	  0 & 0 & -1 \end{array} \right),
\end{equation}
and the matrix $U$, which rotates $X$ into the Cartesian basis, is given by~\cite{Aoki:2007ah}
\begin{equation}
	 U = \left( \begin{array}{ccc}
	- \frac{1}{\sqrt{2}} & 0 &  \frac{1}{\sqrt{2}} \\
	 - \frac{i}{\sqrt{2}} & 0  &   - \frac{i}{\sqrt{2}} \\
	   0 & 1 & 0 \end{array} \right).
	 \label{eq:U}
\end{equation}
In the notation of the parameters of the scalar potential, our chosen benchmark point corresponds to values of
\begin{align}
&\mu_2^2 = -(92.0~\text{GeV})^2, \nn &\lambda_2 = \lambda_3 = {} &\lambda_4 = \lambda_5 =  0.1,\\
&\mu_3^2 = (300~\text{GeV})^2, & M_1 = M_2 &=  100 \text{ GeV.} \nn \\
&\lambda_1 = 0.0468, 
\end{align}
Here $\mu_2^2$ and $\lambda_1$ have respectively been set using $G_F$ and $M_h=125$ GeV [see Eq.~(\ref{eq:lambda1})].

The vevs are obtained by solving the minimization conditions,
\begin{eqnarray}
	&& 
	v_{\phi} \left[ \mu_2^2 + 4 \lambda_1 v_{\phi}^2 
	+ 3 \left( 2 \lambda_2 - \lambda_5 \right) v_{\chi}^2 - \frac{3}{2} M_1 v_{\chi} \right] = 0,
	\nonumber \\
	&&
	3 \mu_3^2 v_{\chi} + 3 \left( 2 \lambda_2 - \lambda_5 \right) v_{\phi}^2 v_{\chi}
	+ 12 \left( \lambda_3 + 3 \lambda_4 \right) v_{\chi}^3
	- \frac{3}{4} M_1 v_{\phi}^2 - 18 M_2 v_{\chi}^2 = 0.
\end{eqnarray}
After electroweak symmetry breaking, the masses of the custodial fiveplet and triplet scalars are respectively given by
\begin{eqnarray}
	m_5^2 &=& \frac{M_1}{4 v_{\chi}} v_\phi^2 + 12 M_2 v_{\chi} 
	+ \frac{3}{2} \lambda_5 v_{\phi}^2 + 8 \lambda_3 v_{\chi}^2, \nonumber \\
	m_3^2 &=&  \frac{M_1}{4 v_{\chi}} (v_\phi^2 + 8 v_{\chi}^2) 
	+ \frac{\lambda_5}{2} (v_{\phi}^2 + 8 v_{\chi}^2) 
	= \left(  \frac{M_1}{4 v_{\chi}} + \frac{\lambda_5}{2} \right) v^2.
\end{eqnarray}

The mixing of the custodial singlets is controlled by the $2\times 2$ mass-squared matrix
\begin{equation}
	\mathcal{M}^2 = \left( \begin{array}{cc}
			\mathcal{M}_{11}^2 & \mathcal{M}_{12}^2 \\
			\mathcal{M}_{12}^2 & \mathcal{M}_{22}^2 \end{array} \right),
\end{equation}
where
\begin{eqnarray}
	\mathcal{M}_{11}^2 &=& 8 \lambda_1 v_{\phi}^2, \nonumber \\
	\mathcal{M}_{12}^2 &=& \frac{\sqrt{3}}{2} v_{\phi} 
	\left[ - M_1 + 4 \left(2 \lambda_2 - \lambda_5 \right) v_{\chi} \right], \nonumber \\
	\mathcal{M}_{22}^2 &=& \frac{M_1 v_{\phi}^2}{4 v_{\chi}} - 6 M_2 v_{\chi} 
	+ 8 \left( \lambda_3 + 3 \lambda_4 \right) v_{\chi}^2.
\end{eqnarray}
The mixing angle $\alpha$ is then fixed by 
\begin{eqnarray}
	\sin 2 \alpha &=&  \frac{2 \mathcal{M}^2_{12}}{m_H^2 - m_h^2},    \nonumber  \\
	\cos 2 \alpha &=&  \frac{ \mathcal{M}^2_{22} - \mathcal{M}^2_{11}  }{m_H^2 - m_h^2},    
\end{eqnarray}
and the singlet masses are given by
\begin{eqnarray}
	m^2_{h,H} &=& \frac{1}{2} \left[ \mathcal{M}_{11}^2 + \mathcal{M}_{22}^2
	\mp \sqrt{\left( \mathcal{M}_{11}^2 - \mathcal{M}_{22}^2 \right)^2 
	+ 4 \left( \mathcal{M}_{12}^2 \right)^2} \right].
	\label{eq:hmass}
\end{eqnarray}

The relationship that allows $\lambda_1$ to be fixed in terms of the measured mass of the observed SM-like Higgs boson is obtained by inverting Eq.~(\ref{eq:hmass}):
\begin{equation}
	\lambda_1 = \frac{1}{8 v_{\phi}^2} \left[ m_h^2 
	+ \frac{\left( \mathcal{M}_{12}^2 \right)^2}{\mathcal{M}_{22}^2 - m_h^2} \right].
	\label{eq:lambda1}
\end{equation}


\end{document}